\documentstyle[aps,twocolumn]{revtex}
\input{epsfig.sty}
\def\coa#1#2#3{{Comput.  Optim.  Appl.} {\bf #1}, #2 (#3).}

\def\science#1#2#3{{Science.} {\bf #1}, #2 (#3).}

\def\nature#1#2#3{{Nature.} {\bf #1}, #2 (#3).}

\def\prl#1#2#3{{Phys.  Rev.  Lett.} {\bf #1}, #2 (#3).}

\def\pre#1#2#3{{Phys.  Rev.  E.} {\bf #1}, #2 (#3).}

\def\jcp#1#2#3{{J. Chem.  Phys.} {\bf #1}, #2 (#3).}
\def\jpc#1#2#3{{J. Phys.  Chem.} {\bf #1}, #2 (#3).}
\def\jcc#1#2#3{{J. Comput.  Chem.} {\bf #1}, #2 (#3).}
\def\jpca#1#2#3{{J. Phys.  Chem.  A.} {\bf #1}, #2 (#3).}

\def\protein#1#2#3{{Proteins} {\bf #1}, #2 (#3).}

\def\N{$N$}

\def\t{{\alpha}}

\def\beq{\begin{equation}}
\def\eqn{\end{equation}}
\def\bc{\begin{center}}
\def\ec{\end{center}}
\begin{document}
\title{Global Optimization on an Evolving Energy Landscape}
\author{J. S. Hunjan, S. Sarkar, and R. Ramaswamy}
\address{School of Physical Sciences\\
Jawaharlal Nehru University.  New Delhi 110 067}
\date{\today}
\maketitle
\begin{abstract}
Locating the global minimum of a complex potential energy surface is
facilitated by considering a homotopy, namely a family of surfaces that
interpolate continuously from an arbitrary initial potential to the system
under consideration.  Different strategies can be used to follow the
evolving minima.  It is possible to enhance the probability of locating the
global minimum through a heuristic choice of interpolation schemes and
parameters, and the continuously evolving potential landscape reduces the
probability of trapping in local minima.  In application to a model
problem, finding the ground--state configuration and energy of
rare--gas (Lennard--Jones) atomic clusters, we demonstrate the utility and
efficacy of this method.

%Apart from identifying the lowest energy configuration, we also show 
%that so--called ``magic'' clusters have anomalously large excitation gaps.
\end{abstract}

{\it Introduction:} Global optimization problems \cite{horst} can often be
formulated in terms of finding the minimum (or maximum) of a
multidimensional potential energy surface (PES).  Such problems, which
occur in a variety of areas, are of considerable practical and theoretical
interest \cite{science2}.  The ``energy landscape'' \cite {walestypes}
paradigm is particularly useful when the potential energy function is
continuously varying with the physical configurations relevant to the
problem .  An example of such a situation is the protein--folding problem
\cite{protein}, namely determining the native configuration of complex
molecules given their atomic composition.  A simpler variant is the
determination of the ground state configuration of atomic or
molecular clusters \cite{doyec}.

In this Letter we propose a new homotopy method to study such problems by
a controlled deformation of the potential energy surface. If $V_f$ is the 
potential energy hypersurface under consideration, we study the landscapes
\beq
\label{eq1}
V(\t) = (1-\t) V_i + \t V_f,
\eqn
with $\t$ a parameter. Given a choice of initial potential, $V_i$,
this is a 1--parameter family of potential energy surfaces which 
smoothly evolves from $V_i$ into $V_f$ as $\t$ varies from $0 \to 1$.

The minima of the landscapes continuously change with $\t$, and in order
to track them, one of two strategies are possible. Varying the 
interpolation parameter $\t$ in a finite number of steps, a
standard technique such as  conjugate gradient (CG) minimization
\cite{cg} can be employed at each $\t$.  On the other hand, one can
consider $\t$ as a time--dependent function such that the
PES evolves according to 
\beq
\label{eq2}
V(t) = (1-h(t)) V_i + h(t) V_f,
\eqn
where $h(t)$ is suitably chosen with $h(0)=0$, and $\lim_{t\to T} h(t) \to
1$.  Over a timescale $T$, therefore, the potential deforms from the
initial to the desired potential energy surface, and the evolving minima
can be tracked, for example, by following the damped dynamics in this
potential via molecular dynamics (MD) simulation.

In the present work we follow both these strategies, and show how homotopic
deformation facilitates location of the global minimum in a model problem. 
Similar (so--called ``continuation'') homotopic methods have frequently
been employed in related situations, as for example in finding roots of
polynomial equations in several variables \cite{homotopy} or in the
mean--field dynamics in attractor neural--networks \cite{amit}.

Different global optimization methods frequently find optimal solutions by
elimination, by seeking lower and lower minima.  Trapping in local
minima---and escape from these minima---is a major practical issue.  A
number of different strategies have been suggested in order to engineer
escape from local minima.  These include both techniques to allow for large
excursions in the phase space by the use of temperature or similar
auxiliary parameters (such as simulated annealing \cite{sa1} and its
variants \cite{lee,dittes}) as well as methods that deform the 
potential energy surface.  The
diffusion equation method \cite{scheraga1} and the distance scaling method
\cite{dsm} fall in this latter class.  Other methods utilize both
strategies, as for example the stochastic tunneling method
\cite{wenzel} where simulated annealing is performed on a surface where
the barriers are exponentially reduced so as to facilitate escape from
local minima, the landscape paving technique \cite{lp},
or the basin hopping technique \cite{walesanddoye} which
replaces the potential surface by a set of piecewise flat regions.

The present technique is in the class of optimization methods that
exploit potential surface deformation to avoid trapping in local minima. 
The interpolation parameter $\alpha$, or the switching functions $h(t)$
smoothly convert one PES into another.  The intermediate potentials are
qualitatively not very different from the asymptotic potential in terms of
the number of minima and maxima, although the relative depths and
curvatures are quite different.  As we discuss below, this feature
contributes to efficiency of the present technique in locating minima. 
The lowest energy achieved when an ensemble of suitably compact initial
configurations is evolved is taken as the ground state
prediction of this method.

{\it Application: } The problem of minimum energy configuration 
determination for $N$ particle atomic clusters is computationally hard,
and the validity of a global solution cannot, typically, be verified. 
Existing data for global minima \cite{ccd} are usually the ``lowest minima
as yet located'' in all but the simplest cases.  
A variety of global optimization techniques have been applied to this
problem \cite{walesanddoye,tsallis} with differing degrees of success.  

For the most extensively studied such systems, namely model rare--gas
clusters, the potential energy surface (PES) is an additive
pairwise Lennard-Jones interaction,
\beq
\label{lj}
V_f = \sum_{i<j} V(r_{ij}) = \sum_{i<j} 4\epsilon [ ({\sigma \over
r_{ij}})^{12} - ({\sigma \over r_{ij}})^{6}]
\eqn
where $r_{ij}$ is the distance between particles $i$ and $j$, and
$\epsilon, \sigma$ are the standard Lennard-Jones parameters.  The
potential energy landscape varies greatly with cluster size.  Notable
difficult optimization problems in this regard are, for example, 38, 75, or
98 atom clusters, where the potential energy surface has the so--called
multiple funnel structure \cite{lj381,doye2,doye}.

In the implementation of the MD approach we proceed as follows. 
$V_i$ is taken to be a pairwise sum
of harmonic terms $V(r_{ij}) = (r_{ij}- 2^{1/6}\sigma)^2/2$.  We perform
molecular dynamics simulations
\cite{verlet} of the $N$ particle system, with an additional damping term
for each particle,
\beq
m\ddot{\vec r_i}+ \gamma \dot{\vec r_i}+
\frac{\partial V(t)}{\partial \vec r_i} = 0 , i=1,\ldots,N
\eqn
where $m$ is the mass of the particle and $\gamma$ is the damping
coefficient.  The internal timescale of interparticle vibrations depends on
the parameters $m, \sigma$ and $\epsilon$.  For a given switching function
$h(t)$ (we have explored a variety of such functions listed in Table I) the
adiabatic time scale is set by the parameter $\zeta$; the entire system
dynamics thus has two external time scales $\zeta^{-1}$ and $m\gamma^{-1}$. 
In the limit $\gamma \to \infty$, our procedure reduces to a steepest
descent minimization on the evolving potential.  The dynamics of the system
is followed until a stationary configuration is reached.

In order to quantitatively assess the efficiency of this procedure, we
define the measure
\beq
{\cal P}_g = \frac{\mbox{Number of~ground state configurations}}
{\mbox{Total~number~of~condensates}},
\eqn
a condensate being a configuration such that all atoms are within a single
cluster.  This is clearly a function of $\gamma$ and
$\zeta$.  For the ground state energy, comparison is made to the existing
benchmark calculations already available for Lennard--Jones clusters
\cite{ccd}.

In the CG approach, $V_i$ is taken to be $\beta \sum_{j=1}^{N}(\vec r_j -
\vec r_j^0)^2$, $\vec r_j^0$ being the (random) initial position for the
$j$th atom.  This choice of $V_i$ ensures that the initial configuration is
the {\it exact global minimum} for the potential energy surface,
Eq.~(\ref{eq1}) with $\alpha = 0$; $\beta$ is a constant that tunes
the curvature of the PES. The parameter $\alpha$ is then varied from 0 to 1 in
$N_s$ discrete steps; the result of the CG minimization (we
follow the standard method \cite{cg}) at each step is taken to be
the starting configuration for the CG minimization at the next value of 
$\alpha$. In this latter approach, therefore, the attempt is to allow 
the global minimum itself to evolve homotopically.

{\it Results:} The present application is intended to be illustrative
rather than exhaustive.  We have systematically studied different cluster
sizes up to $N=40$ and in all cases the calculated ground--state energy and
configuration matches existing results exactly.  This includes the
difficult case of the 38-atom cluster which is an interesting and important
test for any optimization method \cite{lj381}.  The number of minima
increases exponentially with cluster size\cite{stillinger}; for LJ$_7$
there are 4 minima, while for LJ$_{55}$ the number exceeds $10^{10}$
\cite{science2}.  Detailed results, which clarify some aspects of the
present technique are presented for the cases of \N=19,22 (MD) and
\N=38 (CG).

In the MD version of the present technique, in the absence of switching,
namely in the sudden limit $V(t)= V_f$, the system quickly settles into the
nearest available minimum based on the level of damping introduced.  By
starting from an ensemble of initial conditions, a variety of different
minima are reached but the probability of finding the true ground state is
essentially zero for large clusters.  With an adiabatic switch \cite{as}, the
results are dramatically different.  The
continuous evolution of the potential energy landscape is a key factor in
permitting escape from local minima.  Only asymptotically does the system
come to rest, but until then, there is always residual kinetic energy due
to which the system avoids being trapped by small barriers.  
Shown in Fig.~1 is the typical
variation of potential energy (in units of $\epsilon$),
which is nonmonotonic once the adiabatic
switching is incorporated.  Regardless of the actual form of the switching,
more than 85\% of all initially random configurations condense, except in
the case where the switching is applied to the repulsive term of the
potential.  Representative data is given in Table I.

As emphasized, the adiabatic optimization proposed here is heuristic.  
The optimal
choice for the parameters $\gamma, \zeta$ for a given cluster size depend
on a number of features such as the interaction potential parameters and
the inherent time-scales.  By scanning over reasonable values of the
parameters, it is possible to determine regions in parameter space with a
higher than average probability of reaching the ground state.  It also
appears that adiabaticity is crucial since the probability of reaching the
ground state increases substantially with decreasing $\zeta$: ${\cal P}_g$
is shown versus $\zeta$ for the 19--atom case in Fig.~2.

In the CG method of following minima during homotopy, the probability
of reaching the global minimum is enhanced through the modification of
the PES curvature. Since $V_i$ adds a uniform positive curvature
at the intermediate stages it effectively suppresses or eliminates many
barrier. To perform some benchmarking of the advantage this gives,
we present, in Table II, data pertaining to 
finding the global minimum for LJ$_{38}$  comparing the present method and
the basin-hopping technique. The three lowest 
minima are at energies -173.928, -173.252 and -173.134 respectively. In either
method, all particles are initially placed randomly inside a sphere
of radius  $({N\over 4})^{1/3}$. In $V_i$, the parameter $\beta$= 100.
In our implementation of the 
basin-hopping algorithm, coordinate displacements are random in the interval
[-0.3,0.3] and the temperature is taken to be 2. An overall
confining potential of the form 
$V_{c} = \sum_i \exp(20(r_i-a)), a = 1+ ({N\over 4})^{1/3}$ 
was added to prevent dissociation, and a standard Polak--Ribiere algorithm 
was used for the
conjugate--gradient minimization \cite{cg} with tolerance set
between 10$^{-5}$ and 10$^{-7}$.  The average computational effort required
is a product of the number of trials needed in order to get to the ground
state on average and the number of function and derivative calls per trial. 
In our implementation of the algorithms, we find that the reduction in
computational effort in locating the global minimum through the homotopy
method is about 40\%.  The relative efficiencies can, however, vary depending 
on the actual choice of the various adjustable parameters in the two
techniques.
In either the MD or the CG version, configurations that do not reach the
global minimum still typically tend to find the lowest energy states, so
that a by-product of this methodology is a considerably detailed map of the
low excitation regime of the cluster.  This feature, however, is not unique
to the present method.

{\it Summary:} We have presented here a method for global optimization
which relies on the guided evolution of the underlying landscape.  The
methodology for finding minima on this surface can vary, and in the
examples presented here, we have used both the conjugate gradient technique
as well as damped molecular dynamics.  (Dynamics in the landscape has been
incorporated in other techniques, for example in genetic algorithms
\cite{other}.)  As in other methods, apart from the global minimum, we also
obtain a detailed picture of the excitation spectrum.  

Within the context of cluster geometry determination itself, several issues
need to be addressed.  
The adiabatic method can be shown to locate ground states even
when there are bifurcations along the deformation pathway \cite{berryfest}.
Is it possible to design more efficient homotopic
deformations?  What is the role of $V_i$ in controlling the efficiency? 

The application here, though in some ways a model problem, has all the
complications that arise in more general optimization problems.  The
success of this simple technique is therefore encouraging.

\noindent
{\it Acknowledgment:} This research is supported by a grant from
the Department of Science and Technology, India.  
%\newpage

%\newpage
Table~I: Representative results using the MD version of the homotopy method
for LJ$_{19}$ and LJ$_{22}$ as examples of magic and nonmagic clusters.
The parameters are $\zeta = \gamma$ = 0.5, for 1000 trials starting from 
random initial conditions. 
\begin{center}
\begin{tabular} { |c|c|c| } \hline
\multicolumn{1}{|c|}{}
 & \multicolumn{2}{|c|}{${\cal P}_g$}\\
 \cline{2-3} $h(t)$&LJ$_{19}$ & LJ$_{22}$\\
\hline
~~~~~~~~~1 (No Switching) &~~~~~~~0.0~~~&~~~~0.0~~~~~\\ \hline
$1-\exp(-\zeta t)$ &0.0081	&0.0083\\
 \hline 
$\sin{\frac{\pi \zeta t}{2T}}$ & 0.0065	&0.0181\\
 \hline 
$\frac{\zeta t}{T} $&0.0044 &0.0138\\
 \hline 
$(\tanh (\zeta t-10)+1)/2$&0.0124  &0.0176\\
 \hline 
$1-\exp(-\zeta t)\cos^2(3 \zeta	t)$ &0.0111  &0.0041\\
 \hline
\end{tabular}
\end{center}
%\newpage

Table~II: Comparative analysis of the homotopy method and basin--hopping. 
For each method, $N_r$ initial configurations are evolved to find the
global minimum in 100 instances for the LJ$_{38}$ cluster.  $N_j, j= 0,1,2$
are the number of times the lowest three minima are found in the two
methods; the number of function and derivative calls needed (per initial
condition) are also indicated to give an estimate of the computational
effort involved.\\

\begin{center}
\begin{tabular} { |l|c|c|c|c|c|c| } \hline
Optimization&$N_r$&$N_0$&$N_1$&$N_2$&Function&Derivative\\
Method&&&&&Calls&Calls\\	
\hline
Basin&937674&100&239&941&3495&154\\
Hopping&&&&&&\\
\hline
Homotopy &195690&100&4&173&9260&475\\
Method&&&&&&\\
\hline
\end{tabular}
\end{center}

\begin{figure}[!htb]
\centerline{\def\epsfsize#1#2{0.5#1}\epsffile{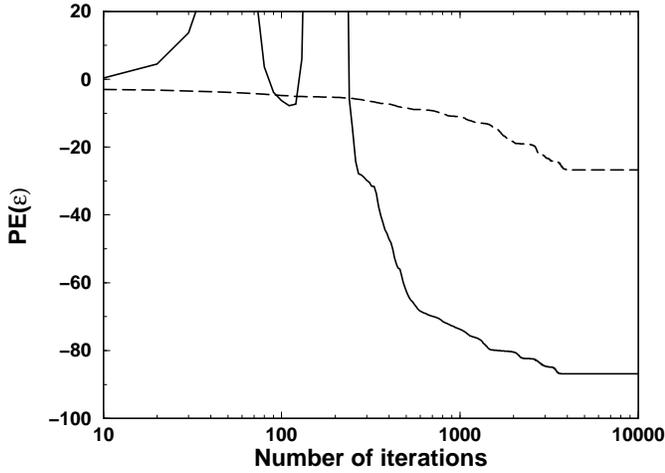}}
\caption
{
\label{fig1.eps}
Typical variation of potential energy (in units of
$\epsilon$) with time for the condensation of
LJ$_{22}$, for the case of no switching, $h(t)=1$ (dashed
line), and with switching (solid line) using $h(t)= 1- \exp(-\zeta t)$.  }
\end{figure}

\begin{figure}[!htb]
\centerline{\def\epsfsize#1#2{0.5#1}\epsffile{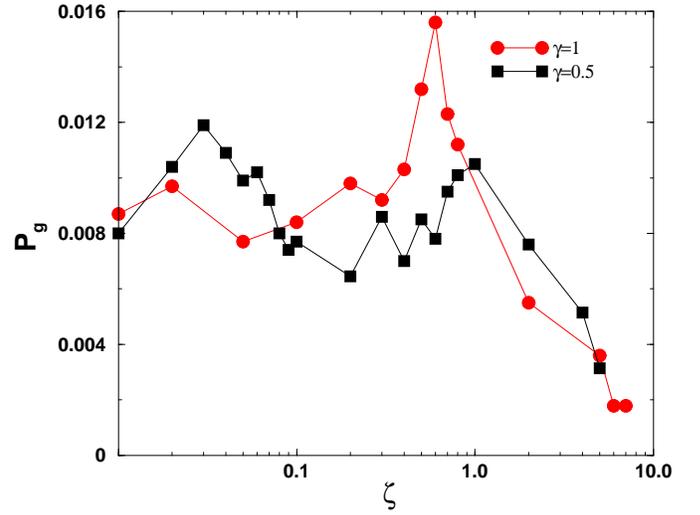}}
\caption
{
Probability of reaching the ground state, ${\cal P}_g$, as a function of
$\zeta$ for $h(t)=1 - \exp -\zeta t$, for the cluster LJ$_{19}$.  }
\end{figure}
\end{document}